# Tunable Superconducting Magnetic Levitation with Self-Stability

Qi Xu, Yi Lin, Yunfei Tan*, and Jianzhao Geng*

Magnetic levitation based on the flux pinning nature of type II superconductors has the merit of self-stability, making it appealing for applications such as high speed bearings, maglev trains, space generators, etc. However, such levitation systems physically rely on the superconductor pre-capturing magnetic flux (i.e. field cooling process) before establishing the levitation state which is nonadjustable afterwards. Moreover, practical type II superconductors in the levitation system inevitably suffer from various sources of energy losses, leading to continuous levitation force decay. These intrinsic drawbacks make superconducting maglev inflexible and impractical for long term operation. Here we propose and demonstrate a new form of superconducting maglev which is tunable and with self-stability. The maglev system uses a closed-loop type II superconducting coil to lock flux of a magnet, establishing self-stable levitation between the two objects. A flux pump is used to modulate the total magnetic flux of the coil without breaking its superconductivity, thus flexibly tuning levitation force and height meanwhile maintaining self-stability. For the first time, we experimentally demonstrate a self-stable type II superconducting maglev system which is able to: counteract long term levitation force decay, adjust levitation force and equilibrium position, and establish levitation under zero field cooling condition. These breakthroughs may bridge the gap between demonstrations and practical applications of type II superconducting maglevs.

Q. Xu, Y. Lin, Prof. Y. Tan, Prof. J. Geng
Wuhan National High Magnetic Field Center
Huazhong University of Science and Technology
Wuhan 430074 P.R. China
E-mail: tanyf@hust.edu.cn, gengjianzhao@hust.edu.cn

# 1. Introduction

Magnetic levitation (maglev), which uses electromagnetic forces to counteract gravity and levitate objects, has been widely used in ultra-high speed rotatory or linear systems to avoid mechanical friction and in static platforms as a magnetic spring to reduce oscillation. Based on different working principles, maglev systems can be mainly classified into four types. Electro-magnetic levitation[1,2] system relies on electromagnetic attraction forces between magnets or magnetic materials, which is not self-stable and requires active control with unavoidable delays. Electro-dynamic levitation[3,4] relies on repulsive forces induced by fast relative motion between a magnet and a conductive object, which is not suitable for static or low speed systems. Type I superconductor-based levitation system[5] relies on the Meissner Effect generated weak repulsive force which only has a single equilibrium point,[6] around which the levitated object spins. In contrast, type II superconductors can pin magnetic flux in the form of vortices,[7,8] and establish levitation with three-dimensional self-stability under field cooling (FC) condition,[9,10] either statically or under motion. Moreover, high-$T_c$ superconducting (HTS) materials are not only able to work at higher temperatures, but also with superior current density under high magnetic fields,[11-15] capable of generating strong levitation forces. These merits make type II superconducting maglev with HTS bulks,[16-20] stacked tapes,[21,22] open loop coil,[23] or closed loop coil[24] promising for applications such as high speed bearings,[25-28] maglev trains,[29-32] space generator,[33] spacecraft docking,[34-37] etc.

Despite the fact that type II superconducting maglev is potentially useful, three intrinsic barriers hinder it from becoming practically usable. Firstly, the levitation force and levitation height inevitably and irreversibly decay.[38-42] This is on one hand due to the flux creep[43] nature of type II superconductors under static levitation, and on the other hand due to dynamic ripple magnet field generated crossed-field demagnetization[44,45] effect during dynamic mechanical oscillations. Therefore, such a maglev system can work as a demonstrator for hours or even days but not feasible for practically long-term operation. Secondly, self-stable levitation can only be established under FC condition.[46-48] Under zero field cooling (ZFC) condition, the equilibrium position locates at infinity if the critical current density of the superconductor is high enough, indicating the levitation is not self-stable or at least with very narrow stable margin. This drawback makes the levitation unsuitable for precisely docking two objects with relative motion, as cooling down the superconductor takes rather long time. It also makes superconductor-superconductor levitation system which normally has much higher levitation force density than superconductor-permanent magnet levitation system complicated to achieve.[49] Last but not least, none of levitation position, levitation force, or levitation stiffness is actively adjustable after levitation state is established under FC condition.[50-52] This intrinsic drawback makes the superconducting levitation unsuitable for applications that require frequently and precisely adjusting levitation states, such as the photolithography platform.

Here, we propose and demonstrate a new form of superconducting magnetic levitation which is called flux-modulated superconducting maglev (FMSM). The FMSM system consists of a permanent magnet (PM), a closed-loop HTS coil, and a rectifier-type flux pump (FP).[53-55] The closed-loop HTS coil has very low resistance, and is able to "lock" the magnetic flux generated by the PM under FC condition, thus establishing self-stable levitation state, just similar to that in existing superconducting maglev systems.

The FP is used to modulate (add or subtract) the total magnetic flux that links the HTS coil without breaking its superconducting state, thus achieving tunable maglev without breaking its self-stability. For the first time, a self-stable type II superconducting maglev system that can counteract levitation height decay, flexibly tune the equilibrium position, and establish levitation under ZFC condition is experimentally demonstrated. This work provides a general methodology towards tunable and self-stable superconducting maglev, which may prompt its wide applications in maglev trains, high speed bearings, maglev platforms, spacecraft docking, and space generators.

## 2. Mechanism and Results

**Figure 1**$a_1$-$a_3$ are the schematic illustration of the proposed FMSM principle. As shown in Figure 1$a_1$, a PM is placed below a closed-loop HTS coil before the coil is cooled to superconducting state, and the PM's magnetic flux links the coil. After the HTS coil is cooled to superconducting state, it will keep the conservation of linked magnetic flux according to the Lenz's law. If the PM is released, it tends to fall due to gravity, inducing a screening current in the HTS coil. The Lorenz force between the screening current and the PM's magnetic field counteracts the PM's gravity, establishing levitation, as shown in Figure 1$a_2$. If the relative position between the HTS coil and the PM changes, the screening current will change automatically. Within a certain range, the varying Lorenz force tends to drag the PM back to the initial equilibrium position, achieving self-stable levitation. In order to tune the levitation state, we can modulate (add or subtract) the total magnetic flux that links coil. This can be realized by unidirectional moving fluxoids across a portion of the HTS coil (we call the superconducting "bridge") using a FP. The process of modulating magnetic flux is equivalent to generating a "transport" current in the closed-loop coil, as shown in Figure 1$a_3$. Both the direction and the value of the transport current can be flexibly modulated, therefore apart from the screening current induced levitation force, additional attractional or repulsive forces can be generated. In this way, adjustable levitation force and levitation height can be achieved. It is important to notice that during flux modulation, the superconducting bridge can stay at a moderate flux flow state in which the equivalent resistance is rather low, and after flux modulation the bridge returns to superconducting immediately. In both cases, the screening current in the HTS coil experiences a very low resistance path, and self-stable levitation can maintain.

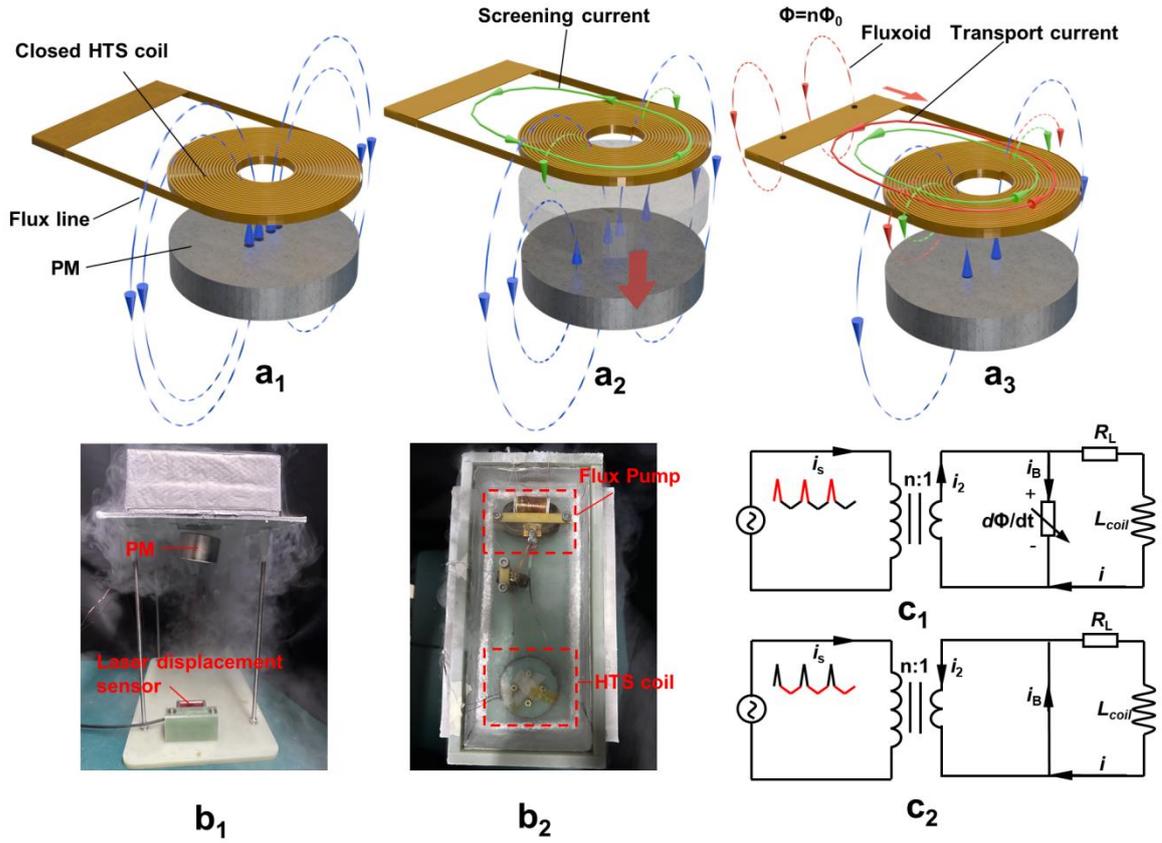

Figure 1. The flux-modulated superconducting maglev (FMSM) principle and experimental system. $a_1$-$a_3$) Schmetic illustration of the FMSM principle. The FMSM uses a closed-loop HTS coil to "lock" the magnetic flux of the PM, using screening current generated Lorentz force to achieve self-stable levitation, and fluxoids are controlled by a flux pump to unidirectionally move across a superconducting portion of the coil, thus generating an equivalent "transport" current. The transport current can be flexible tuned to achieve tunable levitation, meanwhile the screening current ensures the levitation to be self-stable. $b_1$-$b_2$) The front and top view of the FMSM experimental system which consists of a PM, a closed-loop HTS coil, a flux pump, and a laser displacement sensor. $c_1$-$c_2$) Circuit operation of the flux pumped HTS coil system, where during the positive half-cycle of the primary current $i_{s1}$, flux is injected to the HTS coil at a speed of $d\Phi/dt$, and during the negative half-cycle of the primary current the magnetic flux in the HTS coil maintains.

We designed and developed a FMSM demonstration system as shown in Figure $1b_1$-$b_2$. The system consisted of a closed-loop HTS coil (Figure S1, Supporting information), a permanent magnet (Figure S2, Supporting information), a self-rectifier type flux pump (Figure S3, Supporting information) and a laser based displacement sensor. The HTS coil was would from 4 mm wide YBCO coated conductor tape. The terminations of the coil were shorted by soldering a piece of 10 mm wide YBCO coated conductor bridge tape. The self-rectifier consisted of a current-boosting transformer with its secondary winding shorted by the superconducting bridge. Figure $1c_1$-$c_2$ demonstrate the circuit topology and the working principle of the flux pumped HTS coil. The primary winding of the transformer is powered by an asymmetrical alternating current $i_{s1}$, with higher positive peak value than the negative peak value. This current induces a secondary current $i_{s2}$ which has a similar waveform but with much higher amplitudes. Around the positive peak of $i_{s1}$, as shown in Figure $1c_1$, the bridge current $i_b$ slightly exceeds the critical current of the superconductor, thus injecting magnetic flux into the HTS coil at a rate that can be calculated by Equation (1).

$$\frac{d\Phi}{dt} = E_c l \left(\frac{i_b}{I_c}\right)^n \tag{1}$$

Where $\Phi$ is the magnetic flux injected into the HTS coil by flux pump, $E_c$ is a critical electrical field constant and equals to $10^{-4}$ V/m, $n$ depends on the property of the HTS material, $l$ is the length of the HTS bridge, $I_c$ is the critical current of the HTS bridge (Figure S1e, Supporting information). The transport current $i_t$ charged into the HTS coil can be calculated by Equation (2)

$$i_t = \frac{1}{L_{coil}} \int_0^t E_c l \left(\frac{i_b}{I_c}\right)^n dt \tag{2}$$

Around the negative peak of $i_{s1}$, as shown in Figure 1c$_2$, the negative value of $i_b$ is lower than the critical current of the superconducting bridge, the linked magnetic flux of the coil decays slowly due to $R_L$, the sum of joint resistance and flux creep resistance. With the help of the FP, the linked magnetic flux and transport current of the HTS coil can be changed arbitrarily.

In the following, we will demonstrate three key unique functions of the FMSM which are desirable for practical applications but none could be achievable with existing flux pinning based type II superconducting maglev systems, namely: compensating levitation force decay to achieve long-term self-stable levitation at constant height, flexibly adjusting levitation height without breaking self-stability, and establishing self-stable levitation under ZFC condition.

**Figure 2**a$_1$-a$_5$ schematically shows the process of the FMSM to achieve long-term stable levitation with constant height. Initially, once the PM is released after FC, a screening current $i_1$ is induced inside the HTS coil, interacting with the lateral magnetic field to generate a levitation force $F$ on the PM equal to its gravity $G$, thus establishing levitation (Figure 2a$_1$). Due to joint resistance and flux creep in the superconductor, $i_1$ gradually decays to $i_2$, causing $F<G$, resulting in a downward motion trend (Figure 2a$_2$). As shown in Figure 2a$_3$, after the PM falls by a height of $\Delta h$, a larger screening current $i_3$ is induced in the HTS coil, causing $F=G$ again. If the FP is turned on to pump a transport current $i_4$ which has the same direction as $i_3$, it results in $F>G$, and the PM tends to move upwards (Figure 2a$_4$). Assuming that when the PM returns to the initial height in Figure 2a$_1$ with the coincidence $i_5+i_4=i_1$, the final levitation status returns to that in the beginning. It is noted that all above analysis is based on the fact that the motion speed of the PM is low.

We conducted comparative levitation experiments with and without flux modulation under FC condition to verify the analysis. Figure 2b$_1$ demonstrates a cycle of levitation height variation of the FMSM. Initially, the levitation height of the PM was 18 mm (corresponding to Figure 2a$_1$), and after about 80 s of relaxation, it dropped to 19 mm (corresponding to Figure 2a$_3$). Then the FP was actuated, pumping magnetic flux and current into the HTS coil and "dragging" the PM back to 18 mm again (corresponding to Figure 2a$_5$). Figure 2b$_2$ shows the detail of flux modulation and levitation height change. This process was repeated, and the PM was suspended between 18 mm-19 mm for a long time. Figure 2c$_1$-c$_3$ shows the PM levitation height, the HTS coil current, and the HTS coil center field during long term operation respectively. Without flux modulation, the levitation height of the PM continuously decreased over time, while the coil current continuously increased, which is consistent with the explanation in Figure 2a$_1$-a$_3$. The PM finally fell 520 seconds after initial levitation was established (Movie S1, Supporting Information), owing to the fact that the levitation position exceeded the stable margin. For the FMSM, the PM had levitated steadily for 3660 seconds until the experiment was manually ceased (Movie S2, Supporting Information), and the current of the HTS coil stayed constant during the period. It is noted that the stable suspension of the PM exemplifies vertical self-stability of the system. For lateral self-stability verification, we laterally displaced the PM and observed it returning to the equilibrium position, as shown in Movie S3 (Supporting Information). Based on the above experimental results, we can conclude that the FMSM system showed good ability to compensate levitation force and height decay, having achieved long-term levitation with self-stability.

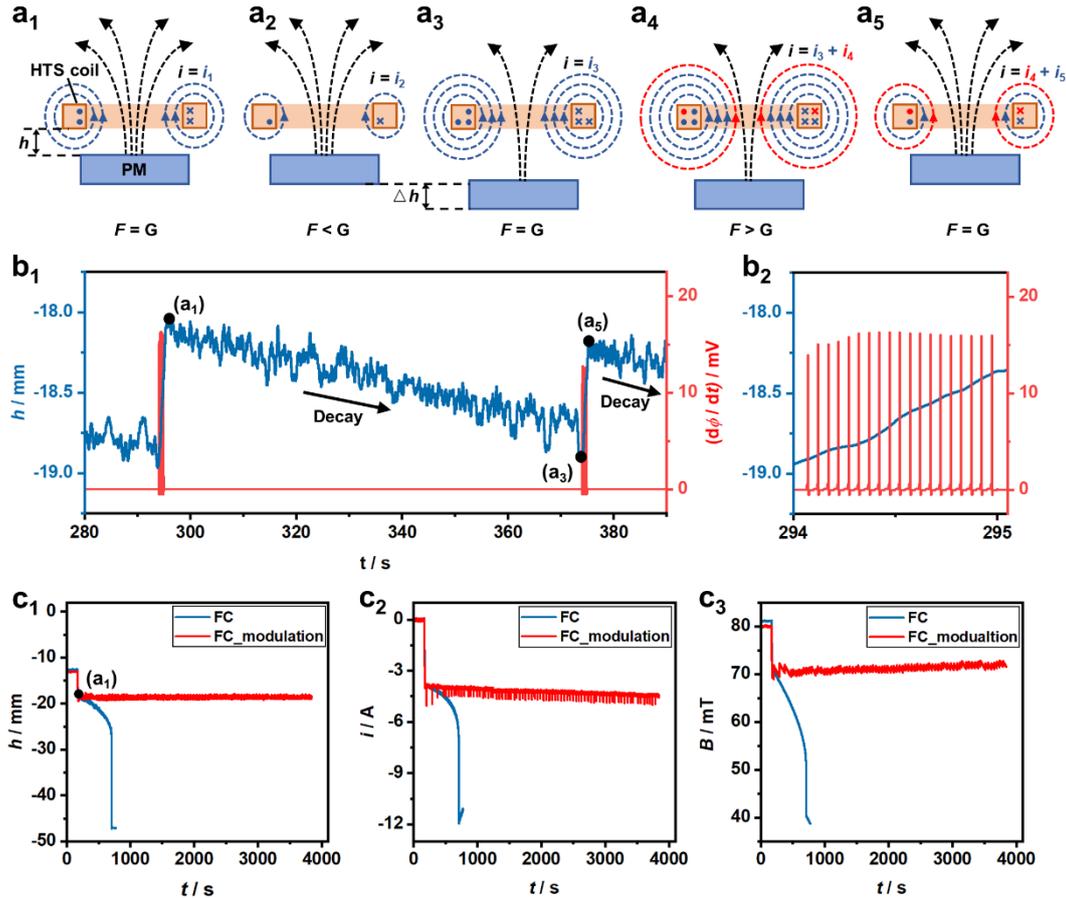

**Figure 2.** Using the FMSM to compensate levitation force decay, and achieve long-term constant-height levitation with three-dimensional self-stability. $a_1$-$a_5$) Schematic illustration of the FMSM compensating the attenuation of levitation height of the PM. $b_1$-$b_2$) Detailed experimental result of the FMSM compensating levitation height decay. $c_1$-$c_3$) Comparison on the PM levitation height, the HTS coil current and its central magnetic field versus time, between with and without flux modulation.

Further, we investigated the ability of the FMSM to modulate the levitation height of the PM, aiming to test whether self-stable levitation could be re-established at positions that largely deviate the initial equilibrium point. As shown in **Figure 3**$a_1$, when the FP continuously charges a transport current in the same direction as the screening current into the HTS coil, the levitation force will increase thus accelerating the PM upward until the PM goes above the initial FC position, then the screening current will reverse, generating a repulsive force. When the sum of the repulsive force and gravity of the PM equals to the transport current generated attractive force, the PM will be suspended stably again. Conversely, as shown in Figure 3$a_2$, when the FP continuously charges a transport current in the opposite direction as the screening current into the HTS coil, the levitation force will decrease thus accelerating the PM downward. The PM would stabilize at a position where the new screening current generated attractive force equals to the transport current generated repulsive force plus the gravity.

Figure 3b shows experimental result on the PM levitation height variation versus time. The initial levitation was established at the height of -25 mm, maintained by flux modulation which counteracted the gradual height relaxation. Then the levitation height was set to change following the sequence of -21 mm, -17 mm, -21 mm, and -25 mm. At each height, the PM was held for 200 s. Experimental details can be found in experimental section. Figure $3c_1$–$c_5$ show snapshot images of the PM at each height and Movie S4 (Supporting Information) recorded the whole changing process of the PM's levitation height. The results show that the FMSM can not only maintain the initial equilibrium point of levitation after FC, but also fast re-establish self-stable levitation at positions that are away from the initial point. This flexibly-tunable, self-stable levitation technique could be potentially useful for static levitation platforms that require frequently and accurately adjusting levitation status.

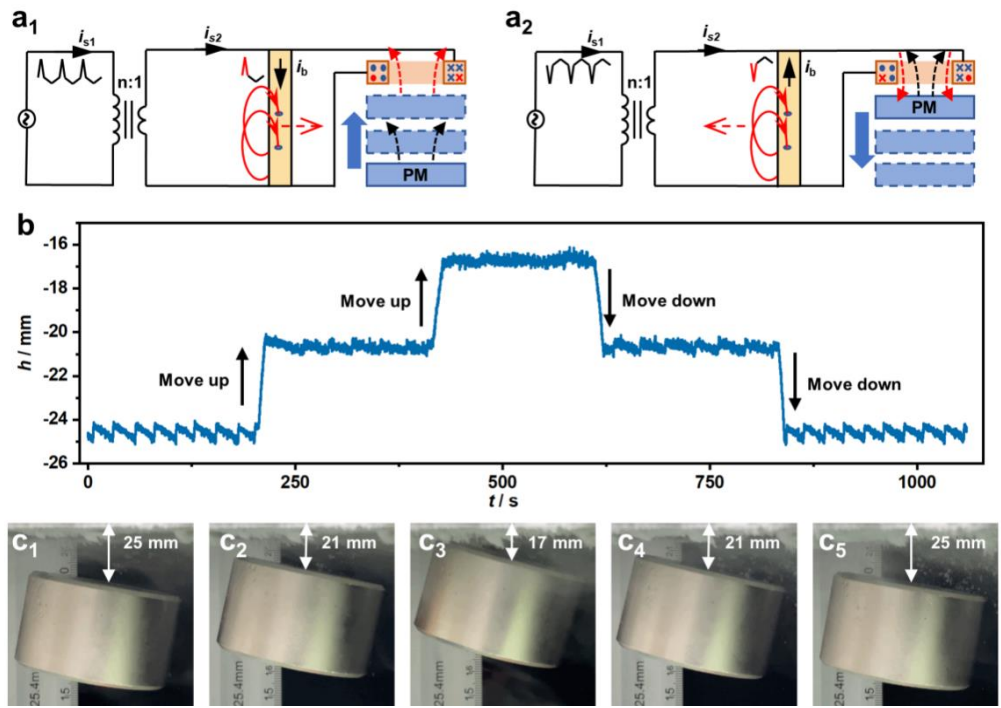

**Figure 3.** Self-stable levitation of the PM with adjustable height. $a_1$-$a_2$) Schematic illustration of the FMSM adjusting levitation height. When the pumped transport current is in the same direction as the screening current, the levitation height of the PM continues to rise, and when the transport current is in the opposite direction of the screening current, the levitation height of the PM continues to fall. b) levitation height modulation result of the FMSM. Three different target levitation heights 25 mm, 21 mm and 17 mm were selected, and the PM was set to levitate stably for 200 s at each height. $c_1$-$c_5$) Optical images captured showing the change of PM's levitation height during the process of flux modulation.

Establishing self-stable levitation under ZFC is rather desirable for many applications but not achievable by conventional flux pinning based type II superconducting maglev. Under ZFC condition, the HTS coil is cooled to superconducting state at a distance far away from the PM. As the PM approaches the HTS coil, an increasing screening current will be induced in the HTS coil to resist its magnetic flux increase. Obviously, the resultant electromagnetic force between the two objects is repulsive, therefore no self-stable levitation could be established, as schematically shown in **Figure 4**$a_1$. This is an intrinsic barrier for all levitation systems using the flux pining property of type II superconductors. But with flux modulation, the situation may be different. Using the FP, a transport current can be generated in the HTS coil with direction opposite to the screening current. If the transport current exactly counteracts the screening current, as shown in Figure 4$a_2$, the electromagnetic status is converted from ZFC to FC. Thus, self-stable levitation can be established after the PM is released, as shown in Figure 4$a_3$. It should be noted that with a pumped transport current either bigger or smaller than the screening current, a self-stable levitation could also be established.

Figure 4b shows finite element modelling results of the electromagnetic force experienced by the PM versus the distance between the PM and the HTS coil under ZFC, with and without flux modulation respectively. The simulation details can be found in Note S1 (Supporting information). The PM was moved upward from 100 mm to 14 mm under the lower surface of the HTS coil, directly downward back to 100 mm after 150 s remaining at 14 mm for the case without flux modulation, while stopped at 14 mm for 150 s until the screening current of the HTS coil being cleared by the FP then downward back to 100 mm for the case with flux modulation. It is noted that in order to realize self-stable levitation of the PM, the electromagnetic force experienced by the PM must have a negative gradient versus distance, and the force should be opposite to and higher than the gravity of the PM which was chosen as 4.5 N (Figure S2b, Supporting information). As shown in Figure 4b, with flux modulation, a segment in the "far away" portion of the force-distance curve satisfies the above conditions, indicating that the PM can be stably suspended below the HTS coil under ZFC process; whereas without flux modulation, that is not achievable.

To further verify the idea, we conducted ZFC levitation experiment with and without flux modulation, respectively. During the experiment, a ball-screw sliding table (Figure S4, Supporting information) vertically lifted the PM from 100 mm to 14 mm under the lower surface of the HTS coil, stopped there for 150 s, and then returned to the initial position, at a speed of 0.5 mm/s. In the flux-modulation case, the FP was turned on to null the induced screening current while the PM was at the 14 mm position. Figure 4$c_1$ shows variation of the height of the PM and the sliding table versus time. We can see that for the no flux-modulation test the PM was always supported by the sliding table indicating no levitation was established, whereas for the test with flux modulation, the PM was stably suspended at 20 mm position as the sliding table being moved downward. The video of both tests can be found in Movie S5 (Supporting information). Figure 4$c_2$ shows the measured HTS coil current $I$ during the above process. In both tests, when the PM was approaching the HTS coil a positive screening current was induced generating repulsive force as above analyzed. Without flux modulation the current gradually decayed when the PM was held at 14 mm position, whereas with flux modulation the current was fast eliminated, thus establishing an electromagnetic status similar to that of the FC process. During the sliding table moving downwards, $I$ in the flux-modulation test directly turned negative generating an attractive force high enough to levitate the PM, in contrast, $I$ in non-flux-modulation test reduced from positive to zero and then turned negative, too small to generate a strong enough levitation force before the PM had been moved far away.

Establishing self-stable maglev under ZFC condition is more flexible and more desirable in many circumstances than under FC condition. A spacecraft docking example is schematically shown in **Figure 5**. When the two crafts approaching each other, it is desirable to magnetically dock them together promptly. Although flux-pinning docking[34-37] interface for spacecraft modules using PM and HTS bulks has been proposed and demonstrated on the ground under FC, it would be rather difficult to be useful in the orbit, since the two modules carrying the PM and the warm superconductor respectively need to be kept at a small and constant distance in the process of the superconductor cooling down, which may take hours in space. Considering the FMSM technique, fast magnetic docking under ZFC may be achieved. As shown in Figure 5b, 5c, when the spacecraft module carrying the pre-cooled HTS coil is approaching the spacecraft module carrying the PM, the FP can eliminate the screening current in real time thus establishing a self-stable magnetic lock. Magnetic docking may even be achieved instantaneously if we have the knowledge of the PM field distribution and pre-pump a transport current into the HTS coil which exactly matches the screening current that could be induced. Another example would be establishing a superconductor-superconductor levitation system which has much higher levitation force than a superconductor-PM system. Under FC condition, the superconductor generating the field must be cooled down prior to the superconductor trapping the field. Thus separate cooling systems are required, largely complicating the technique. If ZFC levitation could be achieved, those two superconductors only need to be cooled down simultaneously, largely simplifying the system. The FMSM could help make ZFC superconductor-superconductor levitation possible, which would be our future work.

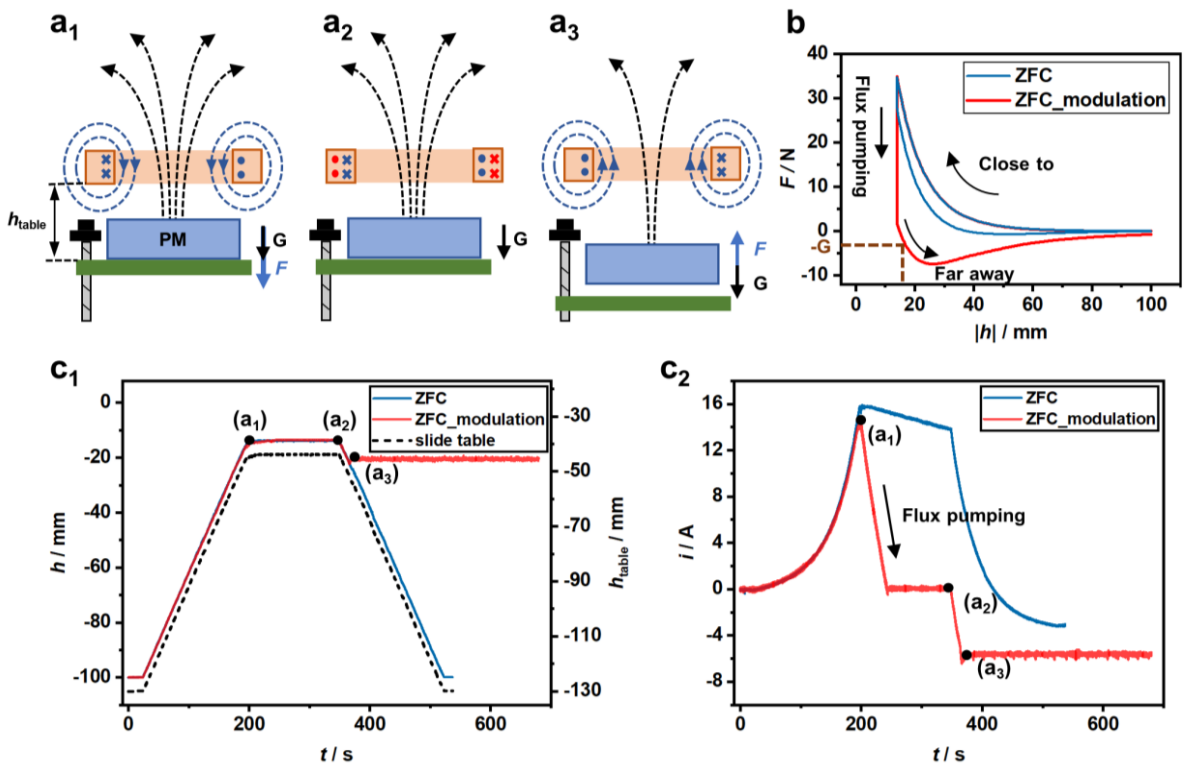

Figure 4. Establishing self-stable levitation under ZFC condition. $a_1$-$a_3$) Schematic illustration of using flux modulation to establish self-stable levitation under ZFC condition. b) the simulated force of PM with respect to the distance between the PM and the HTS coil. Only with flux modulation, self-stable levitation can be achieved. The detailed process of simulation can be found in Note S1 (Supporting information). $c_1$-$c_3$) Experimental results on the variation of the levitation height of the PM and the current of the HTS coil versus time during the ZFC process with and without flux modulation. Without flux modulation, levitation was not established, whereas with flux modulation, the PM stably suspended below the HTS coil.

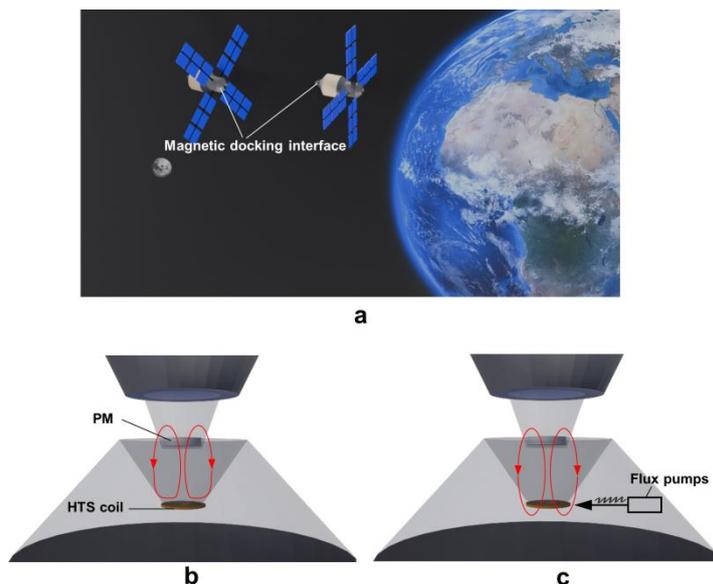

Figure 5. Potential application of the FMSM with ZFC levitation ability in spacecraft docking. a) spacecraft magnetic docking scenario. b) mutual approaching docking interfaces fixed with a pre-cooled HTS coil and a PM before flux modulation. c) with flux modulation in the HTS coil, the two interfaces magnetically docked together.

## 3. Conclusion

This study for the first time proposes and demonstrates a self-stable and tunable superconducting magnetic levitation approach called flux modulated superconducting maglev (FMSM). The self-stability originates from using a closed-loop HTS coil that preserves magnetic flux, similar to flux pinning in type II superconducting bulks. Adjustability is achieved with a flux pump that modulates the total magnetic flux hence the current of the HTS coil without breaking superconductivity. Through experiment, we show three breakthroughs of the FMSM over existing flux pinning based superconducting maglev, i.e. the ability of: counteracting levitation force decay to achieve long-term self-stable levitation, flexibly adjusting levitation height without breaking self-stability, establishing self-stable levitation under zero field cooling condition. These breakthroughs can overcome intrinsic drawbacks of flux pinning based superconducting maglevs, and may bridge the gap for HTS maglevs towards broad applications such as maglev trains, high speed bearings, high precision platforms, spacecraft docking, and space generators.

## 4. Experimental Section

Materials: The YBCO coated conductor tapes of 4 mm and 10 mm width were produced by *Shanghai Superconductor* (Figure S1a, Supporting Information), with a critical temperature <92 K and a critical current of 413 A/cm @ 77 K (Figure S1$e_1$, Supporting Information). The 4 mm wide tape was wound on a 60 mm diameter cylindrical epoxy resin skeleton to form a 120-turns double pancake HTS coil, which has an outer diameter of 65.29 mm shown in Figure S1$b_1$,$b_2$ (Supporting Information). We measured the *E-I* curves of the HTS coil, indicating the coil critical current was 53.9 A at the 100 μV/m criterion which can be discovered in the Figure S1$e_2$ (Supporting Information). Then we soldered the terminations of the HTS coil to a piece of 10 mm wide coated conductor tape (HTS bridge) to form a closed loop (Figure S1c,d, Supporting Information). The FP was utilized to charge the closed-loop HTS coil and then was turned off to measure the current decay curve of the HTS coil, from which the time constant of the HTS coil was calculated to 860 s and the coil inductance was 1.35 mH, joint resistance was 1.7 μΩ (Figure S1$e_3$). The Φ50 mm×30 mm Nd-fe-B PM with 450 g weight has an average surface magnetic intensity of 313.7 mT (Figure S2$c_{1-3}$, Supporting Information). The picture of the self-rectifier type FP was shown in Figure S3a (Supporting Information). The primary winding (Figure S3c, Supporting Information) was wound by 0.5 mm copper wire with 150 turns and had a self-inductance of 15 mH. The secondary winding was a split copper ring shorted by a piece of 10 mm wide YBCO coated conductor tape (Figure S3b, Supporting Information) to form a single turn closed loop. The core of the transformer consists of two U-shaped silicon steels with a saturated magnetic flux density of 1.5 T(Figure S3d, Supporting Information).

Fabrication of the mechanical support of the FMSM experimental system: The FP and the HTS coil were fixed in a liquid nitrogen container assembled from five epoxy plates (Figure S4a, Supporting Information). As shown in Figure S4b (Supporting Information), a steel platform supported by four long screws and an epoxy plate was used to fix the container. It is noted that the all the materials of the platform were not magnetically conductive. The steel plate was designed to be thin (3 mm) to reduce the intrinsic gap between the PM and HTS coil, thus achieving relatively high trapped magntic field. The PM was placed on a ball screw slide table, aligning coaxially with the HTS coil (Figure S4c, Supporting Information). The assembled mechanical support for the system is demonstrated in Figure S4d (Supporting Information).

Control signals generation and data acquisition: We developed a program based on the commercial software *LabVIEW* to control a data acquisition card (DAQ) Ni-6343 (Figure S5, Supporting Information) to generate the excitation signal and to collect the experimental data. As shown in the Figure S5 (Supporting Information), the excitation signal was transferred to a power amplifier (KEPCO BOP50-8DL) to output an asymmetrical current to the primary winding, thus actuating the FP. A hall sensor was placed on the center of the HTS coil to measure its center magnetic field. An open-loop hall sensor was used to measure the current of the HTS coil. A laser based displacement sensor was fixed directly below the PM to measure the height of the PM. The experimental voltage signals from all sensor were collected by NI-6343 and then stored in the computer.

Using the FMSM to achieve long-term constant-height levitation: Initially, the PM was put on the slide table with its top surface 13 mm from the bottom of the HTS coil. Then we poured liquid nitrogen into the container to cool the HTS coil (FC process). Subsequently, the slide table moved down at a speed of 0.5 mm/s until the PM was levitating at 18 mm. Without flux modulation, there was no power supply on the primary windings, thus the levitation height of the PM continously decreased over time. With flux modulation, the program was turned on, sending excitation signal to the power supply once the PM dropped to 19 mm, resulting in transport current in same direction with the screen current pumping in HTS coil with which the PM can be "dragged" back to the initial position.

Self-stable levitation of the PM with adjustable levitation height: In this experiment, we set a target levitation height $h_0$ in the program, if the PM levitation height $h$ is not within the specified range ($h_0-\triangle h \sim h_0+\triangle h$), the program will send the appropriate excitation signal to actuate the FP, achieving the adjustment of levitation height until $h$ change into the specified range. The set target height sequence levitation was -25 mm, -21 mm, -17 mm, -21 mm and -25 mm, and $\triangle h$ was 0.5 mm. At each height, the long-term constant-height levitation was held for 200 s.

Numerical simulation of Maglev force under ZFC condition: The commercial finite element software *Comsol Mutiphysics* was used to conduct 2-D axisymmetric modeling of the HTS coil and FP to simulate the electromagnetic force versus the distance between the PM and the HTS coil under ZFC, with and without flux modulation respectively. The relative motion between the PM and the HTS coil was realized by undirectional coupling of the magnetic field of the PM to the HTS coil boundary in the form of a time-varying boundary condition. The current density and magnetic field distribution of the ZFC process with and without flux pump modulation were studied and contrasted as shown in Figure S7, S8 which demonstrated the current modulation details of the FMSM system. The detailed process of smiulation was described in Note S1 (Supporting Information).

Establishing self-stable levitation under ZFC condition: To conduct the ZFC, firstly, we removed the PM away from the HTS coil using the ball-screw sliding table, stopping at the position where the PM's field had little influence on the HTS coil. After the HTS coil was cooled to become superconducting, the sliding table supported the PM to approch the HTS coil, inducing screening current in the HTS coil associated with strong repulsive electromagnetic force between the PM and the HTS coil. After the PM stopped at the 14 mm position, the FP was triggered to generating a transport current which eleminated the screening current.

## Supporting information

Supporting information is available from the authors.

## Acknowledgement

This work was supported by Nature and Science Foundation of China (grant no.52277010). The authors would like to acknowledge Sizuo Chen and Yunxing Song for experimental support.

## Author Contributions

Q.X. and Y.L. contributed equally to this work. J.G. conceived the idea. J.G. and Y.T. supervised the project. Q.X. and Y.L. conducted the experiment and performed finite element modeling. All authors contributed in writing and revising the manuscript.

## Conflict of Interest

The authors declare no conflict of interest.

## Keywords

magnetic levitation, type II superconductor, HTS coil, flux pump